\documentclass[9pt,twocolumn,twoside]{opticajnl}
\journal{opticajournal}

\setboolean{shortarticle}{true}

\usepackage{lineno}
\usepackage{siunitx}
\usepackage{physics}

\title{High-speed recording technique by synchronous movement of media and spherical reference wave for holographic data storage}

\author[1,2,*]{Shuhei Yoshida}
\author[2]{Atsushi Fukumoto}
\author[2]{Manabu Yamamoto}

\affil[1]{Department of Electrical, Electronic and Communication Engineering, Kindai University, 3-4-1 Kowakae, Higashiosaka City, Osaka 577-8502, Japan}
\affil[2]{HoloStorage Inc., Fujimi Duplex B's 4F, 1-3-11 Fujimi, Chiyoda-ku, Tokyo 102-0071, Japan}

\affil[*]{yshuhei@ele.kindai.ac.jp}

\begin{abstract}
We propose a novel holographic recording technique to improve the recording speed for holographic data storage (HDS). In this technique, holograms are recorded by scanning a digital micromirror device (DMD) that displays a data page with a focused, power density-increased line beam, while synchronously shifting the recording medium and a spherical reference wave. This approach eliminates the stop-and-go motion of driving mechanisms, enabling rapid, continuous, multiplexed recording. Although the recorded localized holograms retain only partial information about the data page, the entire page can be reconstructed simultaneously using the spherical reference wave. Experimental demonstrations achieved a bit error rate (bER) of less than 10\% at a \SI{5}{ms} exposure time and stable multiplexed recording at \SI{150}{Hz}. The proposed technique represents a significant step toward the realization of practical HDS systems.
\end{abstract}

\setboolean{displaycopyright}{false}

\begin{document}

\maketitle

The rapid advancement of information and communication technologies, coupled with the widespread adoption of artificial intelligence (AI), generates unprecedented volumes of data daily. This necessitates storage technologies with significantly enhanced capacities, speeds, and functionalities. Consequently, holographic data storage (HDS)---an optical storage technology based on holography---has attracted considerable research and development interest. HDS theoretically offers large capacities and high transfer rates owing to its three-dimensional (volumetric) recording and two-dimensional, page-based data processing \cite{Psaltis1998,Glass2000,Hesselink2004,Curtis2010a}.

Despite a long history of research \cite{Heerden1963}, practical HDS systems have yet to be realized \cite{Cheriere2025}. While the theoretical potential of HDS is substantial, its optical and mechanical systems remain complex and expensive. Furthermore, practical transfer rate is primarily constrained by the sensitivity of the recording media, laser power, and the operating speeds of spatial light modulators (SLMs), image sensors, and driving mechanisms. Although widely used photopolymer media exhibit higher sensitivity than materials such as photorefractive crystals \cite{Curtis2010b}, watt-class lasers are still required to achieve practical recording rates comparable to existing storage. Currently, widely available photopolymers have sensitivities of approximately $S = \SI{1}{cm/J}$ \cite{Hu2022,Li2022,Jin2024,Hu2024,Peng2025}. Assuming a capacity of 1 Mbit per data page, a recording medium thickness $d$ of \SI{1}{mm}, and a diffraction efficiency $\eta$ of 0.01\% per page, achieving 1 Gbps requires a recording power $I$ of \SI{100}{W/cm^{2}}, based on the definition of sensitivity \cite{Castagna2009,Liu2017}:
\begin{equation}
	S = \frac{1}{Id}\pqty{\pdv{\sqrt{\eta}}{t}}.
\end{equation}
Moreover, the mechanical operation of the translation stage for the recording medium acts as another limiting factor for the recording rate. In particular, the stage's stop-and-go motion induces frequent acceleration and deceleration, drastically reducing the recording rate.

In a shift-multiplexing recording method, multiplexed recording is achieved simply by shifting the recording medium, and this approach has been widely investigated for HDS \cite{Barbastathis1996,Yoshida2013a,Yoshida2013b,Ushiyama2014}. This scheme requires shifting the medium by a specific amount for each hologram recorded. During recording process, a specific exposure time is required due to constraints on laser power and the medium's sensitivity. Consequently, recording cannot be performed while the medium is in motion, and the stage's stop-and-go motion ultimately limits the recording rate. Utilizing high-peak-power pulsed lasers is a potential option. However, HDS requires wavelength-tunable lasers to compensate for Bragg mismatches caused by medium expansion and contraction \cite{Toishi2006,Toishi2007,Tanaka2008,Tanaka2009,Ide2016,Yoshida2018,Qiu2022}; currently, no high-peak-power, wavelength-tunable pulsed lasers are available. Due to these factors, HDS has not yet fully realized its potential, making it difficult to find clear advantages over existing storage technologies.

In this study, we propose a novel holographic recording method to improve the recording speed of HDS. The proposed technique employs a line beam to enhance the signal power per bit, enabling holograms to be recorded in a short time while the medium is in motion. By scanning an SLM with the line beam, a line-shaped signal wave is generated to interfere with a spherical reference wave. Although this line-shaped signal wave contains only a fraction of the data page information, the entire page is recorded in a time-division manner by synchronously shifting the spherical reference wave and the recording medium. This synchronous shift-based recording also eliminates the stop-and-go motion during multiplexing, which has heretofore hindered high-speed recording. While the localized holograms recorded in time-division contain only partial information, the complete data page can be reconstructed by simultaneously illuminating the multiple localized holograms with the spherical reference wave. We experimentally verified the effectiveness of the proposed technique using a custom-built evaluation system, demonstrating hologram recording at several hundred hertz.

\begin{figure}[ht]
	\centering\includegraphics[width=\linewidth]{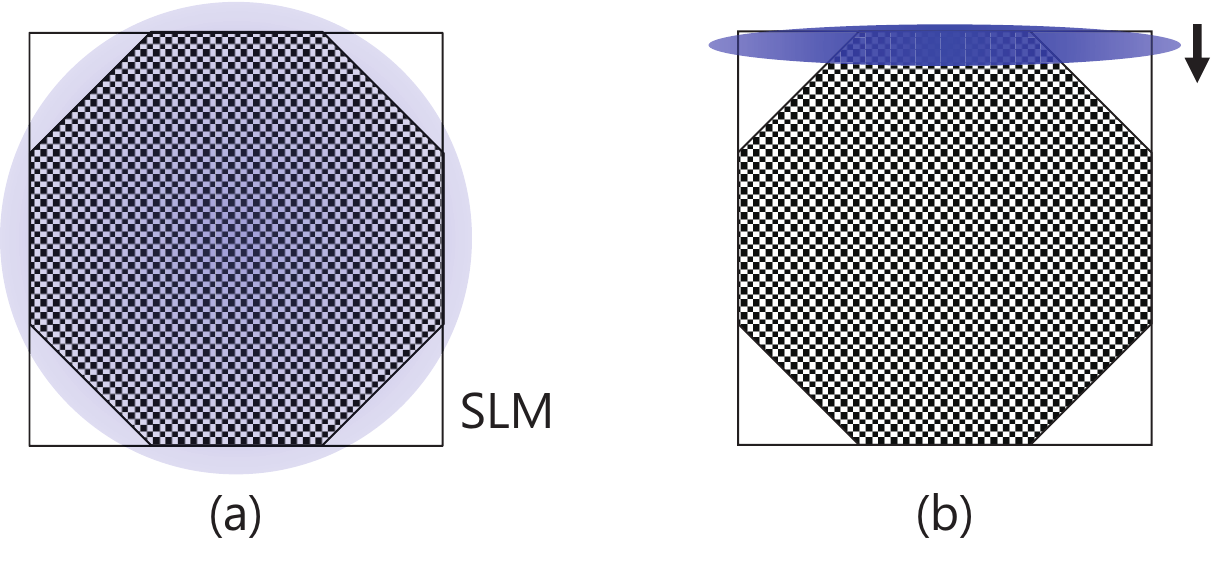}
	\caption{Modulation methods for the signal wave. (a) Conventional scheme. (b) Proposed technique. In the proposed technique, the SLM is scanned with a line beam to record the entire data page in a time-division manner.}
	\label{fig:modulation}
\end{figure}

\begin{figure}[ht]
	\centering\includegraphics[width=\linewidth]{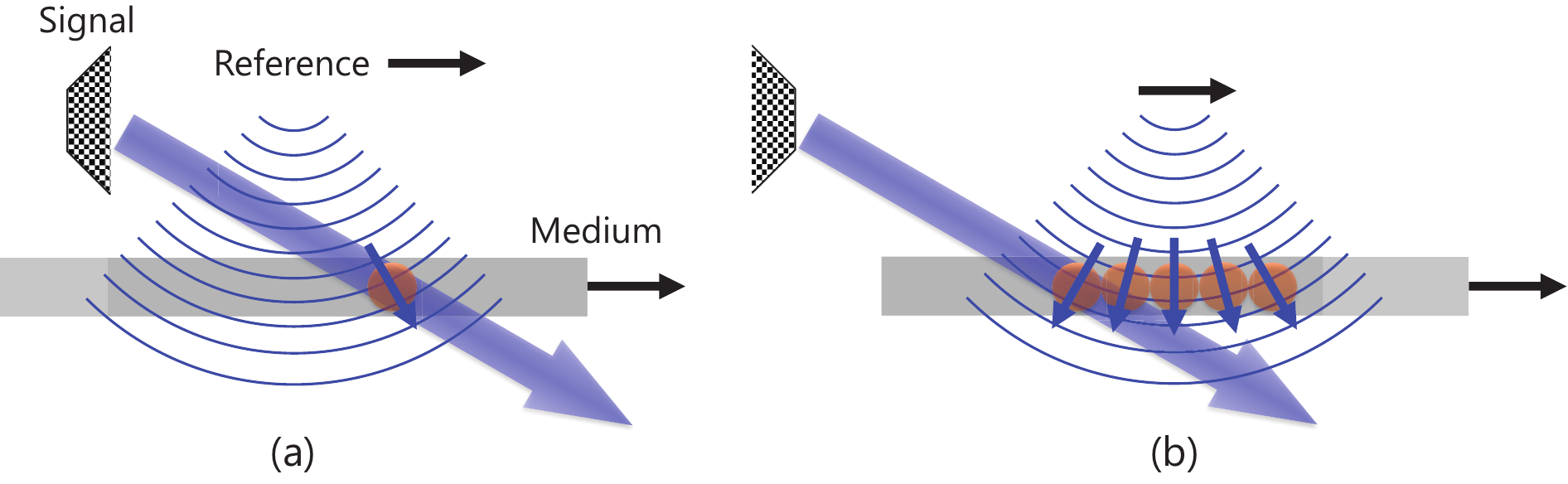}
	\caption{Holographic recording using the proposed technique. The entire data page is recorded in a time-division manner by synchronously shifting the recording medium and the spherical reference wave. Although five holograms are illustrated in the figure, they are actually recorded continuously within the medium.}
	\label{fig:recording}
\end{figure}

\begin{figure}[ht]
	\centering\includegraphics[width=0.5\linewidth]{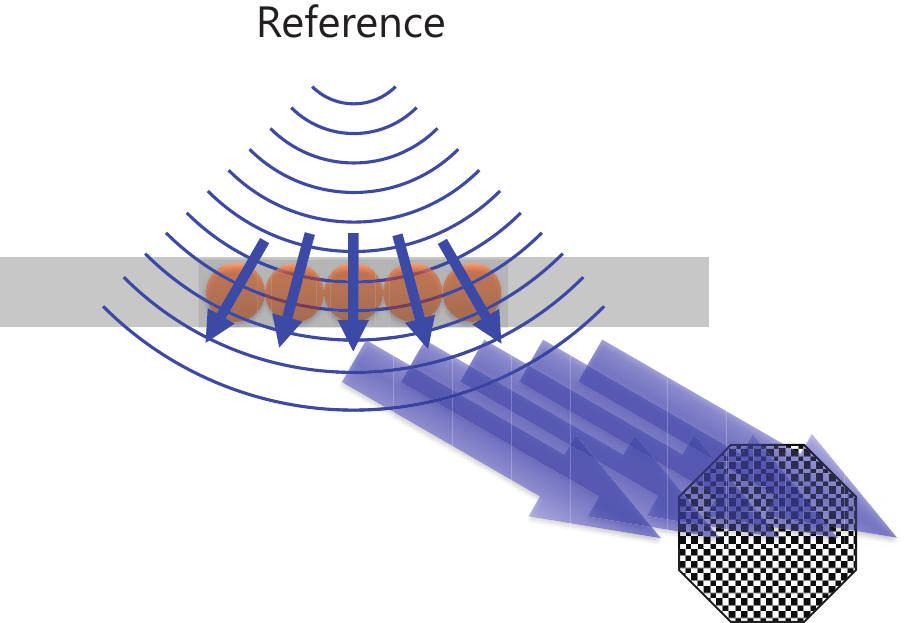}
	\caption{Hologram reconstruction using the proposed technique. Illuminating the center of the continuously recorded holograms with a spherical reference wave satisfies the Bragg condition across the entire recorded area, thereby reconstructing the complete data page.}
	\label{fig:reconstruction}
\end{figure}

In conventional HDS, the beam is expanded to illuminate the SLM as shown in Fig. \ref{fig:modulation}(a). In contrast, the proposed technique scans the SLM using a line beam to enhance the signal power per bit, as illustrated in Fig. \ref{fig:modulation}(b). This line beam interferes with a spherical reference wave on the recording medium, and the resulting interference fringes are recorded as holograms. By shifting the recording medium while synchronously scanning the SLM with the line beam, the entire data page is recorded in a time-division manner.
During hologram recording, the spherical reference wave is shifted synchronously with the recording medium. Because a spherical wave can be considered a superposition of plane waves propagating in different directions, this recording technique enables holograms corresponding to reference waves with various incident angles to be continuously shift-multiplexed on the medium, as illustrated in Fig. \ref{fig:recording}. The proposed technique uses a high-power-per-bit signal, enabling continuous hologram recording while shifting the recording medium, thereby eliminating the stop-and-go motion of the translation stage during multiple recordings. Consequently, the proposed technique achieves high-speed holographic recording by combining increased power density from the line beam with continuous multiplexed recording enabled by synchronous shifting of the recording medium and the reference wave.
During hologram reconstruction, illuminating the recording medium with the spherical reference wave, as shown in Fig. \ref{fig:reconstruction}, ensures that the incident angles of the reference wave match those used during recording for all continuously recorded holograms. Consequently, they are reconstructed simultaneously, allowing the data page recorded in a time-division manner to be retrieved as a single complete page.

Because the proposed technique utilizes a spherical reference wave for both recording and reconstruction, shift-multiplexed recording based on Bragg selectivity is possible \cite{Barbastathis1996,Yoshida2013a,Yoshida2013b,Ushiyama2014}. The shift distance at which the diffraction efficiency of the recorded hologram reaches zero, known as the Bragg null shift, $\Delta x$, is given by:
\begin{equation}
	\Delta x = \frac{\lambda z_{0}}{L \tan\theta_{s}} + \frac{\lambda}{2\mathrm{NA}},
	\label{eq:bragg}
\end{equation}
where, $\lambda$ is the wavelength within the recording medium, $z_{0}$ is the distance from the reference wave source to the medium, $L$ is the thickness of the medium, $\theta_{s}$ is the incident angle of the signal beam, and NA is the numerical aperture of the reference wave. As indicated by Eq. (\ref{eq:bragg}), multiplexed recording with a smaller shift distance can be achieved with a shorter wavelength $\lambda$ and a larger NA.

\begin{figure}[ht]
	\centering\includegraphics[width=\linewidth]{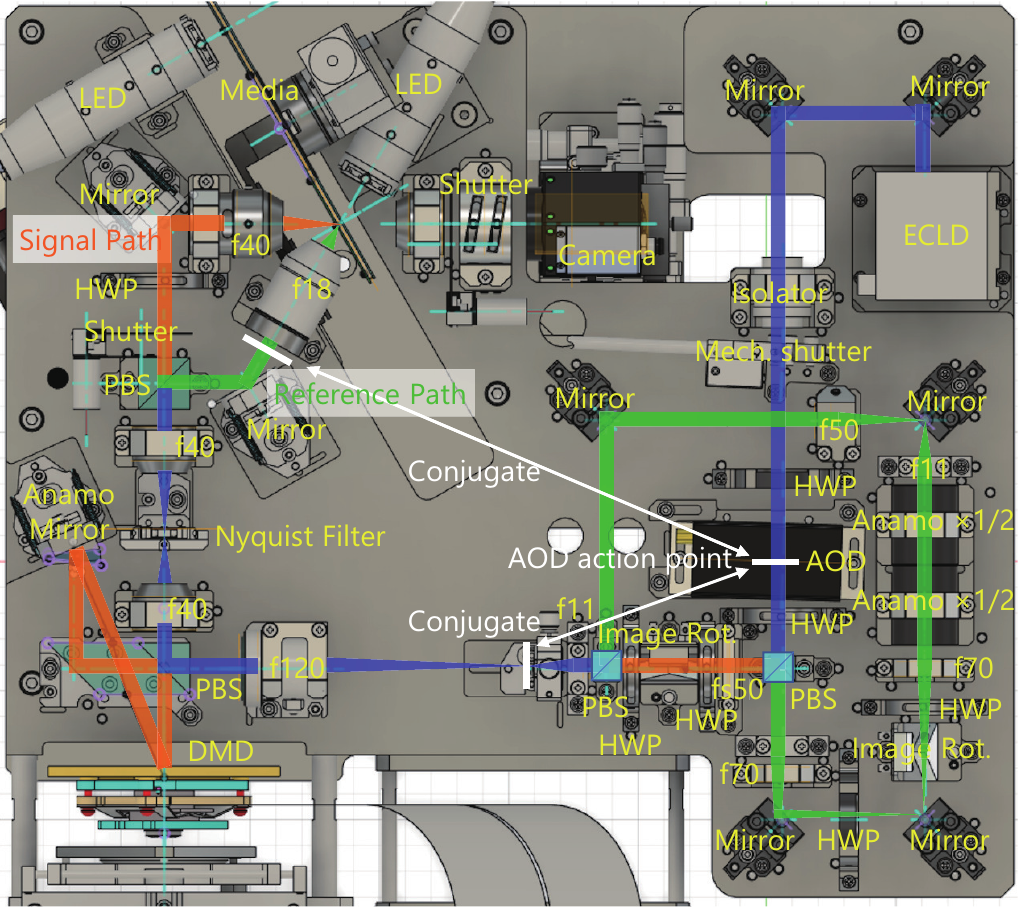}
	\caption{Optical setup for hologram recording and reconstruction. The orange, green, and purple paths represent the signal beam, the reference beam, and the combined signal and reference beams, respectively.}
	\label{fig:optics}
\end{figure}

Figure \ref{fig:optics} illustrates the schematic of the evaluation setup. An external-cavity laser diode (ECLD, Nichia NUV641T) with an output power of \SI{50}{mW} and a central wavelength of \SI{405}{nm} was used as the light source. A digital micromirror device (DMD, ViALUX V-9501) with a resolution of $2560 \times 1600$ and a pixel pitch of \SI{10.8}{\micro m} was employed as the spatial light modulator (SLM). An acousto-optic deflector (AOD, G\&H AODF 4120-3) was utilized to scan the DMD and shift the reference wave. A linear stage (Nippon Pulse SCR075-050) was used to shift the recording medium.
After passing through the AOD, the beam is divided into signal and reference wave paths by a polarizing beam splitter (PBS), relayed by lens systems, and recombined by another PBS. Because the DMD is installed at a \ang{45} angle relative to the optical bench, image rotators are placed in the optical paths of both the signal and reference waves to align the line beam's scanning direction and the reference wave's shifting direction with the DMD's tilt. Initially, the signal and reference waves are s- and p-polarized, respectively; however, they are converted to p- and s-polarization, respectively, by half-wave plates (HWPs) in their respective paths before being recombined at the PBS. The recombined beam is separated again by a PBS after passing through a 4f system. The signal wave passes through three lenses (including a cylindrical lens) from the first PBS to the DMD. Because the front focal point of the third lens is optically conjugate to the AOD, sweeping the beam angle with the AOD allows the line beam to scan the DMD surface. A 4f system relays the signal wave reflected by the DMD, converted to s-polarization (identical to the reference wave) by an HWP, and directed onto the recording medium through an objective lens. Meanwhile, the reference wave is relayed by four 4f systems and then directed onto the recording medium through the objective lens. Because the AOD and the objective lens's front focal point are optically conjugate, sweeping the beam angle with the AOD shifts the reference wave on the recording plane. The designed shift distance of the reference wave on the medium is \SI{6.3}{\micro m}. During hologram reconstruction, a shutter in the signal wave path is closed, and only the recording medium is illuminated by the reference wave. The light diffracted by the hologram (the reconstructed wave) is captured by an image sensor (Baumer VCXU-123M) with a resolution of $4096 \times 3000$ and a pixel pitch of \SI{3.45}{\micro m} to evaluate the signal-to-noise ratio (SNR) and the bit error rate (bER).

\begin{figure}[ht]
	\centering\includegraphics[width=\linewidth]{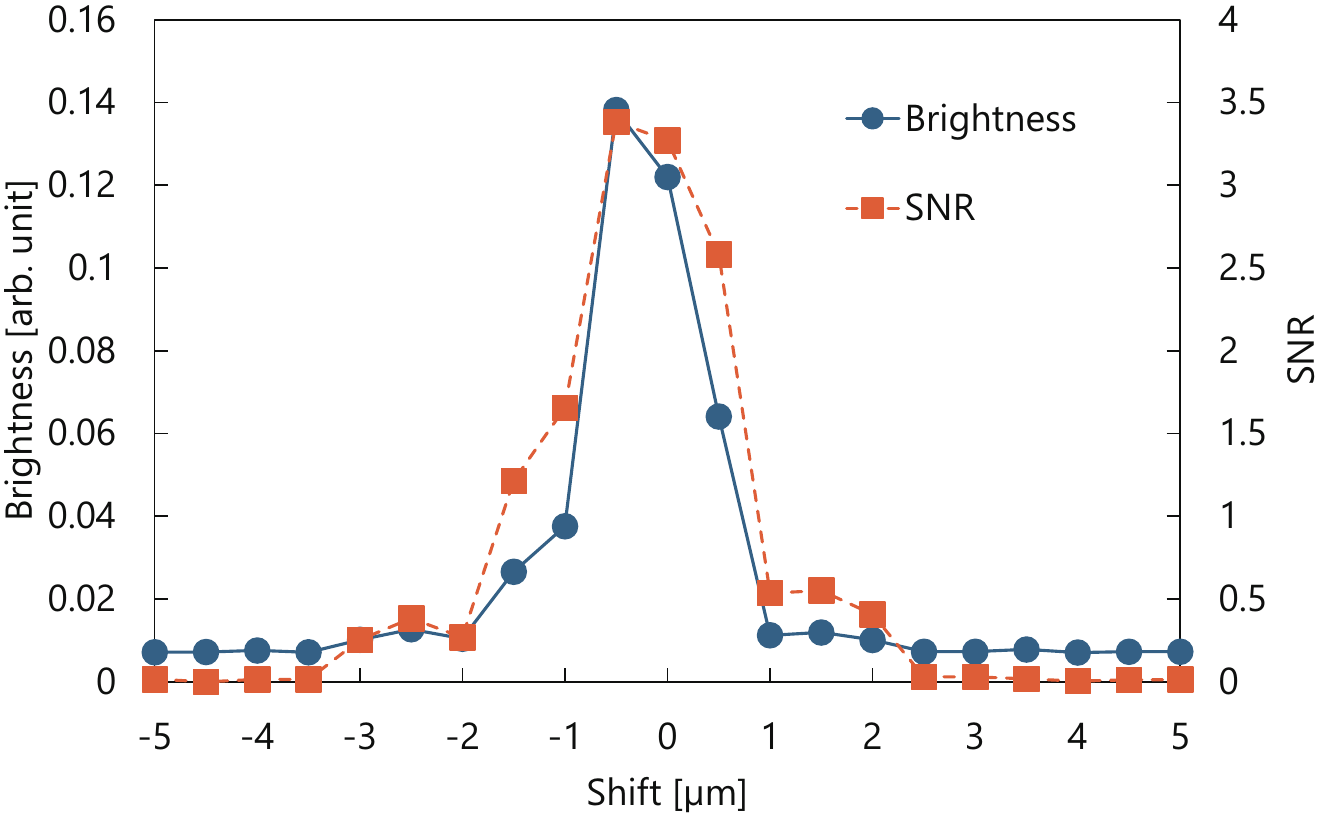}
	\caption{Shift selectivity of a single hologram. The horizontal axis represents the displacement of the reference wave source from its center position in the recorded area, and the vertical axes show the intensity of the reconstructed image (left) and the SNR (right).}
	\label{fig:selectivity}
\end{figure}

Figure \ref{fig:selectivity} shows the evaluation results of the shift selectivity for a single hologram. A \SI{1.5}{mm}-thick photopolymer medium was used in the experiment. The exposure time was set to \SI{6.67}{ms} (equivalent to \SI{150}{Hz}), and recording was performed while driving the linear stage at a speed of \SI{0.750}{mm/s} (calculated as \SI{5}{\micro m}/\SI{6.67}{ms}).
Here, the exposure time refers to the time required to scan the DMD. The shift selectivity was evaluated based on the reconstructed image intensity and the signal-to-noise ratio (SNR). The SNR is defined as:
\begin{equation}
	\mathrm{SNR} = \frac{\mu_{1} - \mu_{0}}{\sqrt{\sigma_{1}^{2} + \sigma_{0}^{2}}},
	\label{eq:snr}
\end{equation}
where $\mu_{0}$ and $\mu_{1}$ are the mean intensity values of the off- and on-pixels, respectively, and $\sigma_{0}^{2}$ and $\sigma_{1}^{2}$ are their respective intensity variances. Although the intensity does not reach zero even with a \SI{5}{\micro m} shift---which is attributed to background noise---the SNR drops to zero at a shift of \SI{5}{\micro m}. These results indicate that shift-multiplexed recording at \SI{5}{\micro m} intervals is feasible in this experimental setup.
\begin{figure}[ht]
	\centering\includegraphics[width=\linewidth]{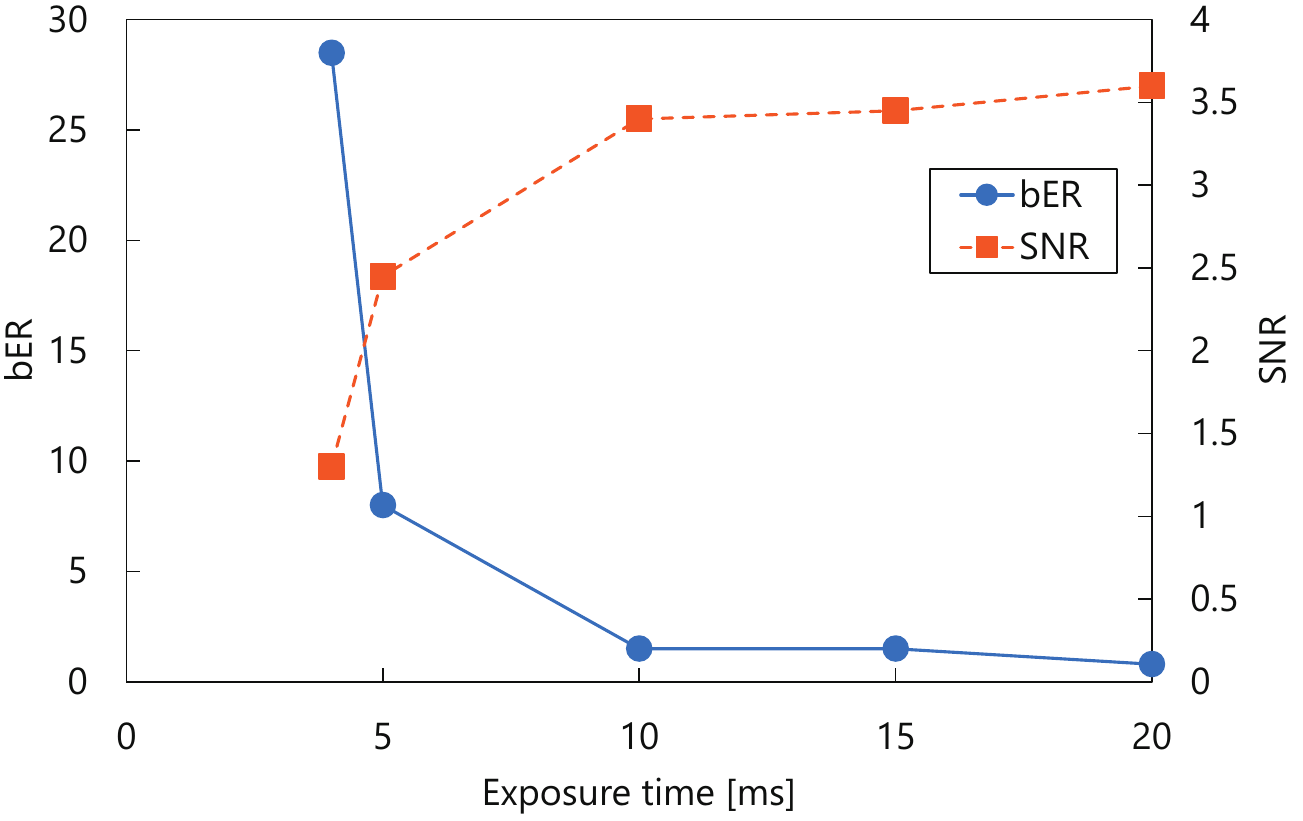}
	\caption{Dependence of bER (left axis) and SNR (right axis) on exposure time (recording speed).}
	\label{fig:ber_snr}
\end{figure}
Figure \ref{fig:ber_snr} illustrates the bER and SNR characteristics as a function of exposure time for single recorded holograms. The driving speed of the linear stage was adjusted to be $\SI{5}{\micro m}/T$ for an exposure time $T$. A bER of less than 10\% was achieved at an exposure time of \SI{5}{ms} (equivalent to \SI{200}{Hz}). These characteristics can be further improved by increasing the laser power.
\begin{figure}[ht]
	\centering\includegraphics[width=\linewidth]{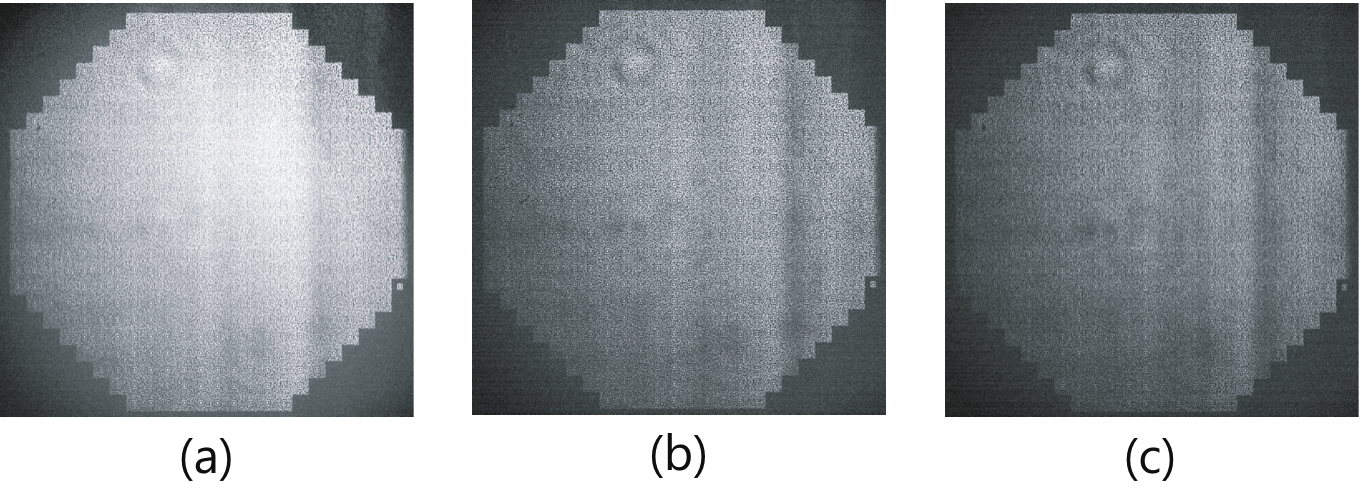}
	\caption{Reconstructed images at different exposure times: (a) \SI{20}{ms}, (b) \SI{10}{ms}, and (c) \SI{5}{ms}.}
	\label{fig:images}
\end{figure}
The corresponding reconstructed images are shown in Fig. \ref{fig:images}. As can be seen, the reconstructed images become darker as the exposure time decreases. This occurs because the refractive index modulation decreases with reduced recording energy, thereby lowering the hologram's diffraction efficiency.

\begin{figure}[ht]
	\centering\includegraphics[width=\linewidth]{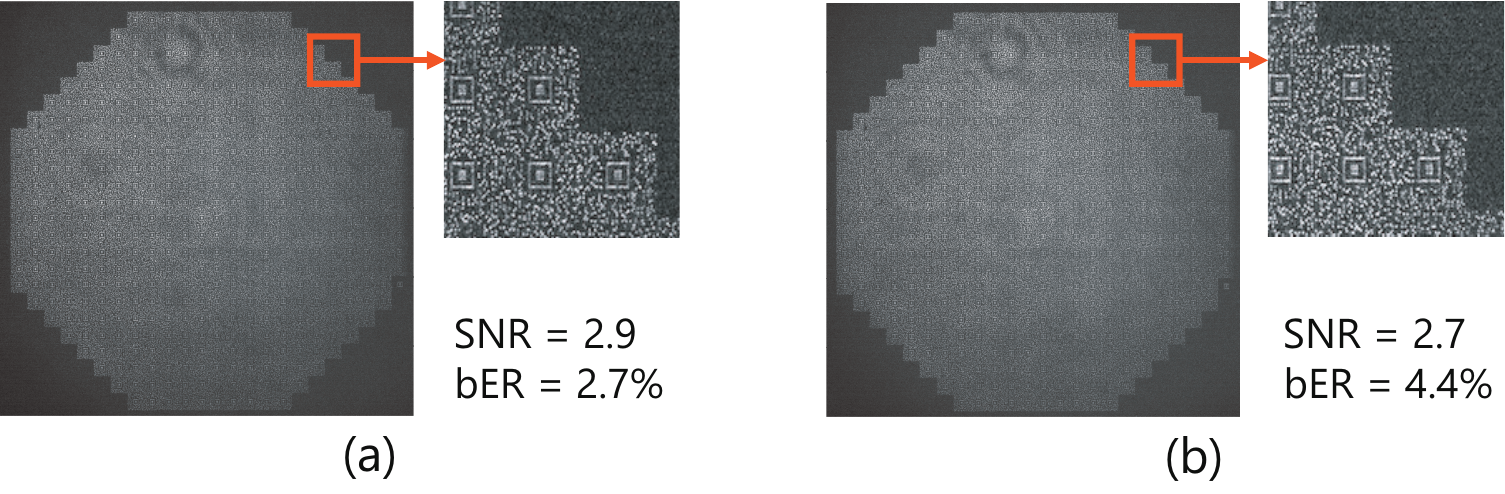}
	\caption{Reconstruction results of multiplexed holograms: reconstructed images of (a) the 5th page out of 11 multiplexed holograms and (b) the 50th page out of 101 multiplexed holograms.}
	\label{fig:multiplexed_images}
\end{figure}

\begin{figure}[ht]
	\centering\includegraphics[width=\linewidth]{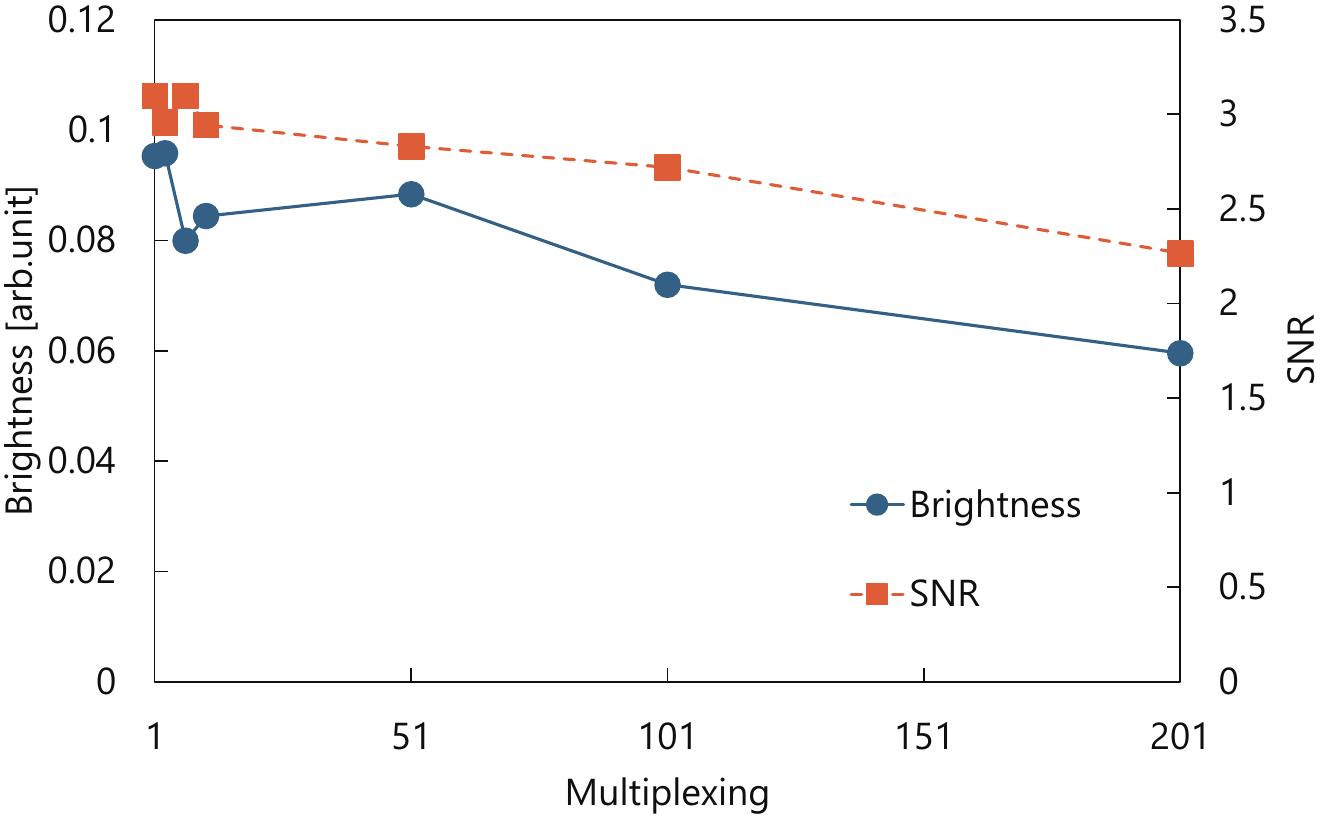}
	\caption{Dependence of intensity (left axis) and SNR (right axis) on the number of multiplexed holograms.}
	\label{fig:intensity_snr}
\end{figure}

Next, we present the evaluation results for multiplexed recording. In the multiplexing experiments, while continuously moving the recording medium, the deflection angle of the AOD was modulated in a sawtooth waveform to repeat the DMD scanning and the reference wave shifting, thereby recording multiple data pages at \SI{5}{\micro m} intervals. Figure \ref{fig:multiplexed_images} shows the reconstruction results of the 5th and 50th data pages when 11 and 101 holograms were multiplexed at \SI{150}{Hz}, respectively. The bER and SNR were 2.7\% and 2.9 for the 5th page, and 4.4\% and 2.7 for the 50th page, respectively. These results demonstrate that stable reconstructed signals can be obtained even during high-speed multiplexed recording.
Figure \ref{fig:intensity_snr} plots the relationship between the number of multiplexed holograms, intensity, and SNR at a recording rate of \SI{150}{Hz}. Both the intensity and the SNR decrease as the number of multiplexed holograms increases. Because this degradation with increasing the number of multiplexed holograms is not caused by inter-page crosstalk, it can be mitigated by increasing the laser power.

In this study, we investigated and experimentally demonstrated a high-speed HDS recording technique that synchronously shifts a spherical reference wave and the recording medium. The proposed technique enhances the signal power per bit by employing a line beam and continuously records holograms in a time-division manner via synchronous shifting of the reference wave and the medium. During reconstruction, the holograms are collectively retrieved using the spherical reference wave. Our experimental evaluations demonstrated that a bER of less than 10\% can be achieved at a recording rate of \SI{200}{Hz}. Furthermore, we achieved stable multiplexed recording at \SI{150}{Hz}, demonstrating a bER of less than 5\%. Further performance enhancements are expected by increasing the laser power. Although practical HDS systems have yet to be fully realized, we are confident that the proposed technique represents a significant step forward toward their realization.

\begin{backmatter}
\bmsection{Disclosures}
The authors declare no conflicts of interest.

\bmsection{Data availability}
Data underlying the results presented in this paper are not publicly available at this time but may be obtained from the authors upon reasonable request.
\end{backmatter}

\bibliography{references}
\bibliographyfullrefs{references}

\end{document}